\documentstyle[preprint,aps,epsf,tighten]{revtex}	
\begin{document}
%
%
\preprint{$
\begin{array}{l}
\mbox{BA-01-20}\\
\mbox{FERMILAB-Pub-01/056-T}\\
\mbox{April 2001}\\[0.3in]
\end{array}
$}
%
%
\title{Realization of the Large Mixing Angle Solar Neutrino Solution\\
	in an SO(10) Supersymmetric Grand Unified Model}

\author{Carl H. Albright}
\address{Department of Physics, Northern Illinois University, DeKalb, IL
60115\\
       and\\
Fermi National Accelerator Laboratory, P.O. Box 500, Batavia, IL
60510\footnote{electronic address: albright@fnal.gov}}
\author{S.M. Barr}
\address{Bartol Research Institute,
University of Delaware, Newark, DE 19716\footnote{electronic address:
smbarr@bartol.udel.edu}}
\maketitle
\begin{abstract}

An $SO(10)$ supersymmetric grand unified model proposed earlier leading to
the solar solution involving ``just-so'' vacuum oscillations is reexamined
to study its ability to obtain the other possible solar solutions.  It is 
found that all four viable solar neutrino oscillation solutions can be 
achieved in the model simply by modification of the right-handed Majorana 
neutrino mass matrix, $M_R$.  Whereas the small mixing and vacuum solutions
are easily obtained with several texture zeros in $M_R$, the currently-favored 
large mixing angle solution requires a nearly geometric hierarchical form for 
$M_R$ that leads by the seesaw formula to a light neutrino mass matrix which 
has two or three texture zeros.  The form of the matrix which provides the 
``fine-tuning'' necessary to achieve the large mixing angle solution 
can be understood in terms of Froggatt-Nielsen diagrams for the Dirac and 
right-handed Majorana neutrino mass matrices.  The solution fulfils several 
leptogenesis requirements which in turn can be responsible for the baryon 
asymmetry in the universe.
\end{abstract}
%
\thispagestyle{empty}
%
%

\section{INTRODUCTION}

Recent results from the Super-Kamiokande Collaboration \cite{atm} involving
atmospheric neutrinos have rather convincingly demonstrated the partial 
disappearance of muon-neutrinos and favor the oscillation of muon-neutrinos 
into tau-neutrinos, rather than into sterile neutrinos at the 99\% 
confidence level.  With regard to solar neutrinos, the situation is 
somewhat more ambiguous.  On the basis of the recently announced 1258 day 
sample results from Super-Kamiokande \cite{solar}, together with the flux 
data from the Chlorine \cite{cl} and Gallium \cite{ga} experiments,
the partial disappearance of electron-neutrinos through 
oscillations into the active flavors of muon- or tau-neutrinos is 
favored over oscillations into purely sterile neutrinos, with the large 
mixing angle (LMA) solution strongly preferred over the small mixing angle
(SMA), the LOW, and the quasi-vacuum (QVO) solutions.  Several 
recent analyses \cite{analyses} based on the smaller 1117 day sample
are basically in agreement with this conclusion by Super-Kamiokande but 
assign slightly higher probabilities to the other three solutions than 
does ref. \cite{solar}.

Whereas the data at present prefer the LMA solution to the solar neutrino 
problem, from a model building point of view the LMA solution seems by far 
the most difficult solution to obtain \cite{smb}.  Many published models of 
neutrino masses and mixings either cannot obtain the LMA solution, or
can only obtain it by fine tuning parameters.  It is thus of importance to
reexamine various approaches to see whether they have sufficient flexibility 
to accomodate the LMA solution in a natural way.

One approach that is particularly flexible is the 
so-called ``lopsided mass matrix" approach. The idea here is that the
large atmospheric neutrino mixing angle arises from the form
of the {\it charged lepton} mass matrix. In other words,
in this approach $U_{\mu 3}$ is more naturally thought of as a mixing
of $\mu$ and $\tau$ rather than of $\nu_{\mu}$ and $\nu_{\tau}$.
On the other hand, the solar neutrino mixing can come from the
{\it neutrino} mass matrix. In this way the atmospheric neutrino problem 
and the solar neutrino problem can be decoupled from each other. This is
one feature that allows the lopsided mass matrix models 
to be more flexible in dealing with the solar neutrino problem.
In this paper we study an especially simple but very predictive example of 
a lopsided mass matrix model to see whether it can accommodate the LMA solution
in a natural way, that is, without fine-tuning. 

The model we shall discuss was developed in a series of papers \cite{abb},
\cite{bimax}, \cite{ab}
by the present authors, together with K.S. Babu earlier in the collaboration.
The model is based on supersymmetric $SO(10)$ grand unification.
As is well known, $SO(10)$ symmetry typically relates the forms
of the Dirac mass matrices of the up quarks, down quarks, charged
leptons and neutrinos (which we denote by $U$, $D$, $L$, and $N$,
respectively) very closely to each other. In this model, 
the lopsidedness of the charged lepton mass matrix, $L$, and of the 
down quark mass matrix, $D$, allow an elegant explanation of many of 
the features of the quark and lepton masses and mixings; in 
particular, the fact that $U_{\mu 3}$ is large whereas $V_{cb}$ is 
small. An interesting point is that in this model the
largeness of the atmospheric mixing angle $U_{\mu 3}$ is forced
upon one by the structure of $L$, which in turn is tied by
$SO(10)$ symmetry to the forms of the other Dirac mass matrices.
On the other hand, as is again typical of $SO(10)$ unification, the
Majorana mass matrix of the right-handed neutrinos, $M_R$, is only
indirectly related to the Dirac matrices, and is therefore much
less constrained. This allows various possibilities for solar neutrino mixing.

In the first papers describing this model \cite{abb}, it was found that the 
SMA solution is very easily obtained if one assumes certain simple
forms for $M_R$, specifically ones which have zeros in the 12, 13, 21 and 
31 elements.  Later it was realized that the QVO solution
is also easily obtained \cite{bimax} by assuming certain other simple forms
for $M_R$. However, it was found that the simplest looking forms for
$M_R$, namely those with many texture zeros, cannot give the LMA
solution \cite{ab}. In light of the recent claim that the LMA solution is
strongly favored, we re-examine this model to see whether the
LMA can be obtained in a natural way. In fact, we look at all four
solar solutions.

In Sect. II we specify the conditions for each of the four solar solutions.
The Dirac mass matrices and parameters obtained earlier for the $SO(10)$ 
model in question are presented in Sect. III, where we also 
numerically determine the structures of the right-handed Majorana matrix
needed to reproduce all four solutions.
A survey of these numerical results in Sect. IV reveals that $M_R$ for 
the LMA solution, in particular, has a remarkably simple texture which can 
be easily related to the Dirac neutrino matrix.  For this 
case, the seesaw mechanism then leads to a light neutrino mass matrix which 
has two or three texture zeros.  The implications of this solution for 
leptogenesis are briefly discussed.

\section{PREFERRED REGIONS IN THE NEUTRINO MIXING PLANE}

Here we summarize the preferred points in the neutrino mixing plane for 
the atmospheric neutrino and the four viable solar neutrino oscillation 
solutions.  We use this information to reconstruct the Maki-Nakagawa-Sakata
(MNS) \cite{MNS} neutrino mixing matrix for each of the four solutions.

For the atmospheric neutrino oscillations, the best fit values obtained 
are \cite{atm}

\begin{equation}
\begin{array}{rcl}
        \Delta m^2_{32} & = & 3.2 \times 10^{-3}\ {\rm eV^2},\\[4pt]
        \sin^2 2\theta_{23} & = & 1.000,\\
\end{array}\label{eq:atm}
\end{equation}
in terms of $\Delta m^2_{ij} \equiv m^2_i - m^2_j$ with $\sin^2 \theta_{atm}
= 4|U_{\mu 3}|^2 |U_{\tau 3}|^2$ expressed in terms of the MNS leptonic 
mixing matrix elements.  Note that to a high degree, the atmospheric 
neutrino mixing is observed to be maximal.  The best fit values for the 
four solar neutrino solutions according to an earlier analysis by 
Gonzalez-Garcia \cite{gg} are
\begin{equation}
\begin{array}{ll}
        SMA: & \Delta m^2_{21} = 5.0 \times 10^{-6}\ {\rm eV^2},\\[2pt]
             & \sin^2 2\theta_{12} = 0.0024,\ \tan^2 \theta_{12} = 
		0.0006,\\[4pt]
        LMA: & \Delta m^2_{21} = 3.2 \times 10^{-5}\ {\rm eV^2},\\[2pt]
             & \sin^2 2\theta_{12} = 0.75,\ \tan^2 \theta_{12} = 0.33,\\[4pt]
        LOW: & \Delta m^2_{21} = 1.0 \times 10^{-7}\ {\rm eV^2},\\[2pt]
             & \sin^2 2\theta_{12} = 0.96,\ \tan^2 \theta_{12} = 0.67,\\[4pt]
        QVO: & \Delta m^2_{21} = 8.6 \times 10^{-10}\ {\rm eV^2},\\[2pt]
             & \sin^2 2\theta_{12} = 0.96,\ \tan^2 \theta_{12} = 1.5.\\[4pt]
\end{array}
\label{eq:mixing}
\end{equation}
In general the MNS mixing matrix, analogous to the CKM quark mixing matrix,
can be written as 
\begin{equation}
  U_{MNS} = \left(\matrix{c_{12}c_{13} & s_{12}c_{13} & s_{13}e^{-i\delta}\cr
        -s_{12}c_{23} - c_{12}s_{23}s_{13}e^{i\delta} & 
                c_{12}c_{23} - s_{12}s_{23}s_{13}e^{i\delta} & s_{23}c_{13}\cr
        s_{12}s_{23} - c_{12}c_{23}s_{13}e^{i\delta} & 
                -c_{12}s_{23} - s_{12}c_{23}s_{13}e^{i\delta} & c_{23}c_{13}
		\cr}\right)
\label{eq:mns}
\end{equation}
in terms of $c_{12} = \cos \theta_{12},\ s_{12} = \sin \theta_{12}$, etc.
With the oscillation parameters relevant to the scenarios indicated above,
to a very good approximation $\theta_{13} = 0^o$ and $\theta_{23} = 45^o$ 
whereby Eq.~(\ref{eq:mns}) becomes essentially real and of the form
\begin{equation}
  U_{MNS} = \left(\matrix{c_{12} & s_{12} & 0\cr 
        -s_{12}/\sqrt{2} & c_{12}/\sqrt{2} & 1/\sqrt{2}\cr
        s_{12}/\sqrt{2} & -c_{12}/\sqrt{2} & 1/\sqrt{2}\cr}\right),
\label{eq:max}
\end{equation}
where the light neutrino mass eigenstates are given in terms of the 
flavor states by 
\begin{equation}
\begin{array}{rcl}
        \nu_3 &=& \frac{1}{\sqrt{2}}(\nu_\mu + \nu_\tau),\\[4pt]
        \nu_2 &=& \nu_e \sin \theta_{12} + \frac{1}{\sqrt{2}}
                (\nu_\mu - \nu_\tau)\cos \theta_{12},\\[4pt]
        \nu_1 &=& \nu_e \cos \theta_{12} - \frac{1}{\sqrt{2}}
                (\nu_\mu - \nu_\tau)\sin \theta_{12}.\\[4pt]
\end{array}
\label{eq:states}
\end{equation}
For the SMA solution, $\theta_{12} = 1.4^o$, while the three large
mixing solar solutions differ from maximal in that the angle is approximately
$30^o$ for the LMA, $39^o$ for the LOW, and $51^o$ for the QVO solutions.
Numerically we find for each case
\begin{equation}
\begin{array}{rl}
  U^{(SMA)}_{MNS}&= \left(\matrix{0.9997 & 0.0241 & 0\cr
	-0.0170 & 0.7069 & 0.7071\cr 0.0170 & -0.7069 & 0.7071\cr}\right),
	\\[0.3in]
  U^{(LMA)}_{MNS}&= \left(\matrix{0.866 & 0.500 & 0\cr
	-0.354 & 0.612 & 0.707\cr 0.354 & -0.612 & 0.707\cr}\right),
	\\[0.3in]
  U^{(LOW)}_{MNS}&= \left(\matrix{0.774 & 0.633 & 0\cr
	-0.448 & 0.547 & 0.707\cr 0.448 & -0.547 & 0.707\cr}\right),
	\\[0.3in]
  U^{(QVO)}_{MNS}&= \left(\matrix{0.633 & 0.775 & 0\cr
	-0.548 & 0.447 & 0.707\cr 0.548 & -0.447 & 0.707\cr}\right).\\
\end{array}
\label{eq:numer}
\end{equation}

\section{MODEL MASS MATRICES AND NUMERICAL DETERMINATIONS OF $M_R$}

The model we are studying here is an $SO(10)$ grand unified model. 
For details of its field content, the flavor symmetry $U(1) \times Z_2 
\times Z_2$, couplings, and so forth, the reader is referred to the
series of papers in which the model was developed \cite{abb}, \cite{ab}. 
Here we will only mention a few of the features of the model important for
the present considerations. 

This model arose from an attempt to construct a realistic $SO(10)$ 
model with the simplest possible, or ``minimal," Higgs content. This 
attempt led very naturally to the following structures at the GUT scale 
for the Dirac mass matrices of the up quarks, down quarks, neutrinos, 
and charged leptons, labeled $U$, $D$, $N$, and $L$, respectively:
\begin{equation}
\begin{array}{ll}
U = \left( \begin{array}{ccc} \eta & 0 & 0 \\
  0 & 0 & \epsilon/3 \\ 0 & - \epsilon/3 & 1 \end{array} \right)M_U,\quad
  & D = \left( \begin{array}{ccc} 0 & \delta & \delta' e^{i\phi}
  \\
  \delta & 0 & \sigma + \epsilon/3  \\
  \delta' e^{i \phi} & - \epsilon/3 & 1 \end{array} \right)M_D, \\ & \\
N = \left( \begin{array}{ccc} \eta & 0 & 0 \\ 0 & 0 & - \epsilon \\
  0 & \epsilon & 1 \end{array} \right)M_U,
  & L = \left( \begin{array}{ccc} 0 & \delta & \delta' e^{i \phi} \\
  \delta & 0 & -\epsilon \\ \delta' e^{i\phi} & 
  \sigma + \epsilon & 1 \end{array} \right)M_D.
\end{array}
\label{eq:Dmatrices}
\end{equation}
A crucial point is that the four Dirac matrices are closely related
to each other by the group theory of $SO(10)$ and
that their forms are definitely fixed in terms of a few parameters.
As a result the model is very predictive, and in fact gives excellent
agreement with all the known facts about the CKM mixings, the quark masses,
and the charged lepton masses. By fitting these data, taking into
account the renormalization effects from the GUT scale to low energies,
the following values of the parameters were obtained:
\begin{equation}
\begin{array}{rlrl}
	M_U&\simeq 113\ {\rm GeV},&\qquad M_D&\simeq 1\ {\rm GeV},\\
	\sigma&=1.78,&\qquad \epsilon&=0.145,\\
	\delta&=0.0086,&\qquad \delta'&= 0.0079,\\
	\phi&= 54^o,&\qquad \eta&= 8 \times 10^{-6}.\\
\end{array}
\label{eq:param}
\end{equation}

A critical feature of the model is that the parameter $\sigma$ is of order
unity, and appears in an asymmetrical or ``lopsided" way in $L$ and $D$.
This fact plays many roles in the model and is indeed the
key to its economy and success in fitting the data. It explains
(a) why $m_c/m_t \ll m_s/m_b$, since $m_c/m_t \sim \epsilon^2$, while
$m_s/m_b \sim \epsilon \sigma$; (b) why the Georgi-Jarlskog relation
$m_s/m_b \cong \frac{1}{3} m_{\mu}/m_{\tau}$ holds, since without
the $\sigma$ term a factor of $\frac{1}{9}$ rather than $\frac{1}{3}$
would result; and (c) why $V_{cb} \ll U_{\mu 3}$. The reason for the
last is that $\sigma$ appears in the 32 element of $L$, where it
causes a large mixing of left-handed muon and tau leptons, i.e. large 
$U_{\mu 3}$, whereas it appears in the 23 element of $D$, where it
causes a large mixing of right-handed quarks, which is not relevant to
$V_{cb}$. The mixing $V_{cb}$ is instead controlled by the 32 element
of $D$, which is the small parameter $\epsilon/3$. The fact that $\sigma$
appears transposed between $D$ and $L$ has to do with the $SU(5)$ structure
of the fields involved.

For present purposes the most important fact is that the largeness of
the atmospheric neutrino mixing angle comes from the parameter $\sigma$
in the charged lepton mass matrix $L$. The contribution of the
neutrino mass matrix to this mixing is formally of order $\epsilon$,
as can be seen from the form of $N$, and is therefore numerically
small for generic choices of $M_R$. On the other hand, one sees that
the solar neutrino mixing angle receives only a small contribution from 
the charged lepton sector, since the 12 and 21 elements of $L$ are
small. Therefore, whether the solar angle is large or small is controlled by
the neutrino mass matrix $M_{\nu} = - N^T M_R^{-1} N$, or in other words by
$M_R$, since $N$ is fixed.
The form of $M_R$ is rather independent of the forms of the
Dirac matrices given in Eq. (\ref{eq:Dmatrices}) because it comes from 
completely different operators. That is why in this
model --- and indeed in the general framework \cite{smb} of ``lopsided mass 
matrix models" in which the atmospheric angle arises from large lopsided
entries in $L$ --- there is great flexibility in how the solar
neutrino problem is solved. Different solar oscillation solutions
can be obtained by changing the form of $M_R$ without affecting
in any way the fits to the CKM parameters, the masses of the quarks
and charged leptons, or the fact that the atmospheric neutrino angle 
is large. 

In our first papers where this model was discussed, forms of $M_R$
were assumed in which the SMA solar solution was naturally obtained.
Indeed, one sees immediately that if $M_R$ has vanishing 12, 21, 13, and 31
elements, $M_{\nu}$ does not contribute to the solar neutrino
angle, which then comes entirely from $L$ and is therefore small.

The QVO solution can also be very easily obtained. In \cite{ab}
the following simple form of $M_R$ was constructed:
\begin{equation}
	M_R = \left(\matrix{ 0 & A\epsilon^3 & 0 \cr A\epsilon^3 & 0 & 0 \cr 
	        0 & 0 & 1 \cr}\right)\Lambda_R.
\label{eq:maj}
\end{equation}
With this form the seesaw formula \cite{gm-r-s} gives the light neutrino mass 
matrix to be
\begin{equation}
  M_\nu = N^T M_R^{-1} N = \left(\matrix{ 0 & 0 & -\eta/(A\epsilon^2)\cr
        0 & \epsilon^2 & \epsilon\cr 
        -\eta/(A\epsilon^2) & \epsilon & 1\cr}\right)M^2_U/\Lambda_R.
\label{eq:lightnu}
\end{equation}
With $\Lambda_R = 2.4 \times 10^{14}$ GeV and $A = 0.05$, a fairly reasonable
fit to the quasi-vacuum solution then emerged with
\begin{equation}
\begin{array}{ll}
  \multicolumn{2}{c}{m_3 = 54.3\ {\rm meV},\ m_2 = 59.6\ {\rm \mu eV},
        \ m_1 = 56.5\ {\rm \mu eV,}}\\[0.1in]
  \multicolumn{2}{c}{M_3 = 2.4 \times 10^{14}\ {\rm GeV},\ M_2 = M_1 =
	3.66 \times 10^{10}\ {\rm GeV,}}\\[0.1in]
  \multicolumn{2}{c}{U_{e2} = 0.733,\ U_{e3} = 0.047,\ U_{\mu 3} = -0.818,
        \ \delta'_{CP} = -0.2^o,}\\[0.1in]
        \Delta m^2_{32} = 3.0 \times 10^{-3}\ {\rm eV^2},\quad &        
                \sin^2 2\theta_{atm} = 0.89,\\[0.1in]
        \Delta m^2_{21} = 3.6 \times 10^{-10}\ {\rm eV^2},\quad &
                \sin^2 2\theta_{solar} = 0.99,\\[0.1in]
        \Delta m^2_{31} = 3.0 \times 10^{-3}\ {\rm eV^2},\quad &
                \sin^2 2\theta_{reac} = 0.009.\\
        \end{array}\\
\end{equation}

We now wish to search for right-handed Majorana mass matrix textures which 
fit more accurately each of the four solar neutrino solutions.
We first note that the MNS mixing matrix corresponds to the product of 
two unitary transformations,
\begin{equation}
	U_{MNS} = U^\dagger _L U_\nu,
\label{eq:Uprod}
\end{equation}
where $U_L$ diagonalizes the Hermitian lepton matrix $L^\dagger L$, and
$U_\nu$ diagonalizes the light neutrino mass matrix which we assume to be
real and symmetric for simplicity:
\begin{equation}
	L^{{\rm diag}\dagger} L^{\rm diag} = U^\dagger _L L^\dagger L U_L,  
	\quad M^{\rm diag}_\nu = U^T_\nu M_\nu U_\nu.
\label{eq:diag}
\end{equation}
It is easy to see that, given a specific pattern of neutrino masses and
mixings, one can invert to find a form of $M_R$ that will give that pattern.
To be given a pattern of neutrino masses and mixings means that one
is given the MNS mixing matrix $U_{MNS}$ and the neutrino mass
eigenvalues $m_1$, $m_2$, and $m_3$. On the other hand, the model itself
specifies the charged lepton matrix, $L$, and the neutrino Dirac mass
matrix, $N$; cf. Eq. (\ref{eq:Dmatrices}). Thus $M_R$ can be inferred as 
follows. First, $U_L$ can be 
directly obtained from diagonalization of $L^{\dag} L$. Then $U_L$ together 
with the given $U_{MNS}$ determine $U_{\nu}$ through Eq. (\ref{eq:Uprod}). 
Although $L$ and hence $U_L$ are complex, we can obtain a real $U_{\nu}$ 
by making use of the freedom to perform a phase rotation on $U_{MNS}$, 
so that 
\begin{equation}
	U_\nu = U_L \cdot diag(1,1,e^{-i\phi}) \cdot U_{MNS}.
\label{eq:Unu}
\end{equation}
Then, defining 
\begin{equation}
	M^{\rm diag}_\nu = diag(m_1,\ -m_2,\ m_3),
\label{eq:Mnudiag}
\end{equation}
with hierarchical masses chosen which are related to the $\Delta m^2_{ij}$'s,
one can use this and the matrix $U_{\nu}$ already found to determine 
$M_{\nu}$ by using the second of Eq. (\ref{eq:diag}). Finally,
one can use the $N$ known from the model and $M_{\nu}$ to find $M_R$ by 
inverting the see-saw formula
\begin{equation}
	M_R = N M^{-1} _\nu N^T.
\label{eq:MR}
\end{equation}

We present the numerical results for each of the four solar solutions as 
follows:
\begin{equation}
\begin{array}{rl}
  M^{(SMA)}_R&= \left(\matrix{0.156 \times 10^{-7} & -0.190 \times 10^{-4} &
		0.116 \times 10^{-3}\cr
	  -0.190 \times 10^{-4} & 0.0105 & -0.123\cr 
	  0.116 \times 10^{-3} & -0.123 & 1.000\cr}\right)\times 5.2 \times 
		10^{14}\ {\rm GeV},\\[0.3in]
	&\ {\rm with\ } M_1 = 3.7 \times 10^6,\ M_2 = 2.3 \times 10^{12},\ 
	   M_3 = 5.3 \times 10^{14}\ {\rm GeV};\\[0.2in]
  M^{(LMA)}_R&= \left(\matrix{8.30 \times 10^{-10} & -0.511 \times 10^{-5} &
		2.13 \times 10^{-5}\cr
	  -0.511 \times 10^{-5} & 0.0244 & -0.155\cr 
	  2.13 \times 10^{-5} & -0.155 & 1.000\cr}\right)\times 3.0 
		\times 10^{14}\ {\rm GeV},\\[0.3in]
	&\ {\rm with\ } M_1 = 4.2 \times 10^6,\ M_2 = 6.7 \times 10^{10},\ 
	   M_3 = 3.1 \times 10^{14}\ {\rm GeV};\\[0.2in]
  M^{(LOW)}_R&= \left(\matrix{5.15 \times 10^{-10} & -1.43 \times 10^{-5} &
		5.46 \times 10^{-5}\cr
	  -1.43 \times 10^{-5} & 0.0292 & -0.176\cr 
	  5.46 \times 10^{-5} & -0.176 & 1.000\cr}\right)\times 5.8 
		\times 10^{14}\ {\rm GeV},\\[0.3in]
	&\ {\rm with\ } M_1 = 6.0 \times 10^6,\ M_2 = 9.7 \times 10^{11},\ 
	   M_3 = 6.0 \times 10^{14}\ {\rm GeV};\\[0.2in]
  M^{(QVO)}_R&= \left(\matrix{-6.98 \times 10^{-10} & -1.33 \times 10^{-5} &
		4.75 \times 10^{-5}\cr
	  -1.33 \times 10^{-5} & 0.0481 & -0.222\cr 
	  4.75 \times 10^{-5} & -0.222 & 1.000\cr}\right)\times 3.2 
		\times 10^{15}\ {\rm GeV},\\[0.3in]
	&\ {\rm with\ } M_1 = 8.8 \times 10^6,\ M_2 = 3.8 \times 10^{12},\ 
	   M_3 = 3.3 \times 10^{15}\ {\rm GeV}.\\
\end{array}
\label{eq:MRnumer}
\end{equation}
Strictly speaking, the above results were obtained at the GUT scale, but 
with the moderate value of $\tan \beta \sim 5$ preferred by the model 
\cite{ab} and for the hierarchical and sign choices given in Eq. 
(\ref{eq:Mnudiag}) above, the evolutions in masses and mixings from the GUT 
scale to the low scales are extremely small and can be neglected \cite{evol}.

That one can find forms for $M_R$ that reproduce the various
solar neutrino solutions is in itself not very significant,
for as we have just seen, this is guaranteed as long as the
relevant matrices are invertible. The significant question is whether
the matrix $M_R$ that gives a certain solar solution is obtainable
in the model under discussion in a simple way without fine-tuning.
The forms for $M_R^{(SMA)}$ and $M_R^{(QVO)}$ given in Eq. (\ref{eq:MRnumer})
are complicated-looking. However, these are the forms that reproduce the
present {\it best-fit} SMA and QVO solutions according to \cite{gg}.  One 
already knows from our previous work, as has already been mentioned, that 
much simpler forms for $M_R$, having several
texture zeros, give perfectly satisfactory SMA and QVO solutions;
moreover, those simpler forms are obtainable straightforwardly
without fine-tuning. But that same earlier work shows that forms
for $M_R$ having several texture zeros do not yield a satisfactory
LMA solution in this model. The question is then whether the
form $M_R^{(LMA)}$ given in Eq. (\ref{eq:MRnumer}), or something sufficiently
close to it, can be obtained simply and naturally in the model.
To this question we now turn.

\section{SIMPLE ANALYTIC FORM FOR $M_R$ INVOLVING THE LMA SOLUTION}

At first glance the form of $M_R^{(LMA)}$ in Eq. (\ref{eq:MRnumer}) looks
very complicated. However, it has some significant features that
suggest that it may be obtainable in a simple way.
First of all, one sees that $(M_R)_{23} = (M_R)_{32} \simeq - \epsilon$
and $(M_R)_{22} \simeq \epsilon^2$, where $\epsilon$ is the
parameter that appears in the Dirac matrix, $N$, of Eq. (\ref{eq:Dmatrices}).
To a good approximation we can therefore introduce the analytic form
\begin{equation}
    M^{(LMA)}_R = \left(\matrix{c^2\eta^2 & -b\epsilon\eta & a\eta\cr
		-b\epsilon\eta & \epsilon^2 & -\epsilon\cr
		a\eta & -\epsilon & 1\cr}\right) \Lambda_R,	
\label{eq:MRapprox}
\end{equation}
written in terms of parameters appearing in the Dirac neutrino matrix, where 
$\epsilon = 0.145$ and $\eta = 0.8\times 10^{-5}$ as before, cf. Eq. 
(\ref{eq:param}), and 
$\Lambda_R = 2.5 \times 10^{14}$ GeV.  It will turn out that the new
parameters $a,\ b$ and $c$ are all of order unity in order to obtain the 
LMA solar solution.  Making use of the seesaw formula, we then find 
\begin{equation}
  M^{(LMA)}_\nu \sim \left(\matrix{0 & \epsilon/(a - b) & 0\cr
	\epsilon/(a - b) & -\epsilon^2 (c^2 - b^2)/(a - b)^2 
	& -b\epsilon/(a - b) \cr
	0 & -b\epsilon/(a - b) & 1\cr}\right) M^2_U /\Lambda_R.
\label{eq:approxlightnu}
\end{equation}

It is interesting, and we shall see, relevant to leptogenesis that this 
form has some texture zeros.  These texture zeros follow directly from the 
form of the 23 block of Eq. (\ref{eq:MRapprox}).
That this 23 block has rank 1 immediately suggests 
that it can arise from diagrams of the Froggatt-Nielson type \cite{f-n}. 
Moreover, the fact that the same parameter $\epsilon$ appears in both 
$M_R$ and $N$ suggests the possibility that the hierarchies in the 23 blocks 
of both matrices may have the same origin. These suggestions can be realized 
as we now show.

In Fig. 1 we repeat for clarity the diagrams in our model which contributed 
to the Dirac matrices in the 2-3 sector.
The dominant 33 elements arise from the ${\bf 10}_H$ Higgs electroweak
doublet contributions.  For the 23 and/or 32 elements, higher-order 
contributions arise from electroweak doublets in both the ${\bf 10}_H$ 
and ${\bf 16}_H$ $SO(10)$ representations, with additional singlet Higgs 
VEV's and a ${\bf 45}_H$ Higgs GUT scale VEV pointing in the $B - L$ 
direction.  Due to the $SU(5)$ structure of the Higgs fields, the diagram 
appearing in Fig. 1(c) contributes only to the $D_{23}$ and $L_{32}$ elements 
of the down quark and charged lepton mass matrices.  Note that the internal 
superheavy fermions appearing in ${\bf 16},\ \overline{\bf 16},\ 
{\bf 10}_1$ and ${\bf 10}_2$ are integrated out.  

In Fig. 2 we show the lower-order diagrams which can contribute to the 
2-3 sector of the right-handed Majorana mass matrix.  Here a singlet 
Higgs GUT scale VEV, $V_M$,  couples two superheavy conjugate singlet 
fermions thus inducing a breaking of lepton number.  The VEV's in the 
$\overline{\bf 16}_H$'s also appear at the GUT scale.  The ${\bf 
\overline{16}_H - 1''_H}$ pair appearing in insertion ``A'' of Fig. 2 serves
to lower the heaviest right-handed Majorana neutrino mass down to 
$\Lambda_R = 2.5 \times 10^{14}$ GeV from the GUT scale value of 
$2 \times 10^{16}$ GeV.  By making use of the techniques spelled out
in detail in \cite{ab}, one can readily show that the 23 elements
of $N$ and $M_R$ are scaled by the same factor $\epsilon$ relative to their 
33 elements.  The factor enters antisymmetrically in $N$ for the 23 and 
32 elements due to the $B - L$ nature of the ${\bf 45}_H$ VEV and the 
presence of both left-handed neutrino and conjugate neutrino states, while
it appears symmetrically in $M_R$ since both states involve conjugate 
neutrinos.  In the Majorana case, both superheavy singlet and ${\bf 45}$
fermions must be integrated out.  We have checked that the these diagrams
can be achieved as indicated with proper assignment of the $U(1) \times
Z_2 \times Z_2$ flavor quantum numbers for the new heavy 
fermion fields introduced.

We now turn to the small entries of the first row and column of $M_R^{(LMA)}$ 
in Eq. (\ref{eq:MRapprox}) with $a$, $b$, and $c$ numbers of order 
unity. The fact that the whole matrix manifests a geometrical hierarchy 
involving the same small parameters $\epsilon$ and $\eta$ that appear in $N$
reinforces the idea that $M_R$ may be simply obtained by Froggatt-Nielsen-type
diagrams involving some of the same VEVs that generate $N$. If it were the 
case that $a = b = c$ exactly, then the whole matrix would have rank 1, and 
thus all its elements could be obtained from a single Yukawa vertex 
$({\bf 1}^c_3 {\bf 1}^c_3)V_M$, in the same way that 
we illustrated for the 23 block. However, that would, of course, 
be unrealistic in that two neutrinos would then be massless.
However, it is not necessary that the matrix be of rank 1 in order
that it arise from simple Froggatt-Nielson diagrams. Thus we have the 
possibility that $a$, $b$, and $c$ are not all equal. 

For an especially interesting numerical example, suppose that 
\begin{equation}
a = 1,\ b = c = 2,\ \Lambda = 2.5 \times 10^{14} {\rm GeV}.
\label{eq:params}
\end{equation}
This has a simple interpretation in that all elements of the $M_R$ matrix
receive contributions from the Yukawa vertex involving $V_M$, while only the
13 and 31 elements receive contributions from a second $\Delta L = 2$ violating
Yukawa vertex involving $V'_M$.  This can be realized with the proper 
choice of flavor indices for $V'_M$.  By the see-saw formula, one
then has
\begin{equation}
	M^{(LMA)}_\nu = \left(\matrix{0 & -\epsilon & 0\cr
		-\epsilon & 0 & 2\epsilon\cr
		0 & 2\epsilon & 1\cr}\right)M^2_U/\Lambda_R 
\label{eq:Mnuspecial}
\end{equation}
with three texture zeros from which we obtain 
\begin{equation} 
\begin{array}{ll}
	\multicolumn{2}{c}{m_3 = 57.4\ {\rm meV},\ m_2 = 9.83\ {\rm meV},\ 
		m_1 = 5.61\ {\rm meV},}\\[0.1in]
	\multicolumn{2}{c}{M_3 = 2.5 \times 10^{14}\ {\rm GeV},\ M_2 = M_1 =
		2.8 \times 10^{8}\ {\rm GeV},}\\[0.1in]
	\multicolumn{2}{c}{U_{e2} = 0.572,\ U_{e3} = -0.014,\ U_{\mu 3} = 
		0.733,\ \delta_{CP} = 0^o,}\\[0.1in]
	\Delta m^2_{32} = 3.2 \times 10^{-3}\ {\rm eV^2},\quad & 
        	\sin^2 2\theta_{atm} = 0.994,\\[0.1in]
	\Delta m^2_{21} = 6.5 \times 10^{-5}\ {\rm eV^2},\quad &
		\sin^2 2\theta_{solar} = 0.88,\\[0.1in]
	\Delta m^2_{31} = 3.2 \times 10^{-3}\ {\rm eV^2},\quad &
		\sin^2 2\theta_{reac} = 0.0008.\\
\end{array}
\label{eq:MRnum}
\end{equation}
These results fit both the atmospheric and the LMA solar mixing solutions 
extremely well and can be considered a success for the model.  In fact, 
the best fit point for the LMA solar mixing
solution as given by the Super-Kamiokande Collaboration in their latest
analysis of 1258 days of data taking \cite{solar} is $(\sin^2 2\theta_{sol} 
= 0.87,\ \Delta m^2_{21} = 7 \times 10^{-5}\ {\rm eV^2})$.  We find the whole 
newly-allowed LMA region can be covered with $a,\ b$ and $c$ varying by 
factors of $O(1)$ from the values given in Eq. (\ref{eq:params}).  It is 
noteworthy that the solar neutrino mixing is near maximal, but not actually 
maximal as that is presently excluded experimentally by the SuperKamiokande 
results at more than the 95\% confidence level.

How fine-tuned is the form of $M_R$ that we have been discussing?
One feature that at least appears fine-tuned is the fact that the
23 and 32 entries in Eq. (\ref{eq:MRapprox}) are not only of order $\epsilon$
but actually equal to $- \epsilon$ exactly. This one has no right
to expect from the mere fact that the same VEVs come into the diagrams
for $N$ and $M_R$, since as can be seen from Figs. 1 and 2 different
Yukawa couplings are involved in the 23 entries of the two matrices.
One can test how fine-tuned the form in Eq. (\ref{eq:MRapprox}) is by replacing
the 23 and 32 elements by $- d \epsilon$ and the 22 element by 
$d^2 \epsilon^2$. (The fact that the same $d$ enters is due to
the factorized structure of the diagrams in Fig. 2, and is therefore
not a fine-tuning.) One naturally expects that $d$ is of order unity,
but how close must it be to 1 to give a realistic LMA solution?
It turns out that the most severe constraint on the value of $d$ comes
from the limit on $U_{e3}$. To satisfy the condition that
$|U_{e3}| \leq 0.15$ \cite{CHOOZ}, one requires that $0.85 \leq d \leq 1.15$.
Thus, the LMA solution does not require an unnatural fine-tuning
of parameters. 

Finally we note that the upper bound on the lightest heavy Majorana 
neutrino mass $M_1$ should be less than or of order $10^9$ GeV to prevent 
overproduction of gravitinos from overclosing the universe after inflation
\cite{lgen}.
This bound is satisfied for all four solar solutions as determined in
Eq. (\ref{eq:MRnumer}) and, in particular, for the model illustrated above.
A second condition for 
leptogenesis is that the 13 and 31 elements of $M_R$ be suppressed by a 
factor of at least $10^3$ relative to the 33 element to inhibit mixing of 
the heaviest right-handed neutrino with the lightest one in order to prevent
its rapid decay washing out the lepton asymmetry generated.  This is 
satisfied in our model.

\section{SUMMARY}

We have investigated how an $SO(10)$ SUSY GUT model proposed earlier can 
be modified in order to obtain solar neutrino solutions other than the 
vacuum solution.  The study revealed that only the right-handed Majorana 
neutrino mass matrix needed to be modified, with the Dirac matrices for 
the neutrinos and charged leptons (as well as for the quarks) left 
unchanged.  In short, in this model the maximal 
atmospheric neutrino mixing is controlled primarily by the structure of the 
charged lepton mass matrix, while the type of solar neutrino solution is 
largely determined by the form of the right-handed Majorana mass matrix.

Of particular interest was the finding that the large mixing angle
solar solution is readily obtained with a nearly geometrical hierarchy in 
$M_R$, where the 2-3 subsector has a close relationship with that for 
the Dirac neutrino matrix, as seen by study of the Froggatt-Nielsen 
diagrams.  It is precisely this structure which provides
the ``fine-tuning'' necessary to achieve the LMA solar solution.\\[1in]

The research of SMB was supported in part by Department of Energy Grant
Number DE FG02 91 ER 40626 A007.  One of us (CHA) thanks the Fermilab
Theoretical Physics Department for its kind hospitality where much of his
work was carried out.  Fermilab is operated by Universities Research 
Association Inc. under contract with the Department of Energy.\\[1in]
%
%

%
%
%
\begin{figure}
\vspace*{0.5in}
\centerline{
\epsfxsize=1\hsize
\epsfbox{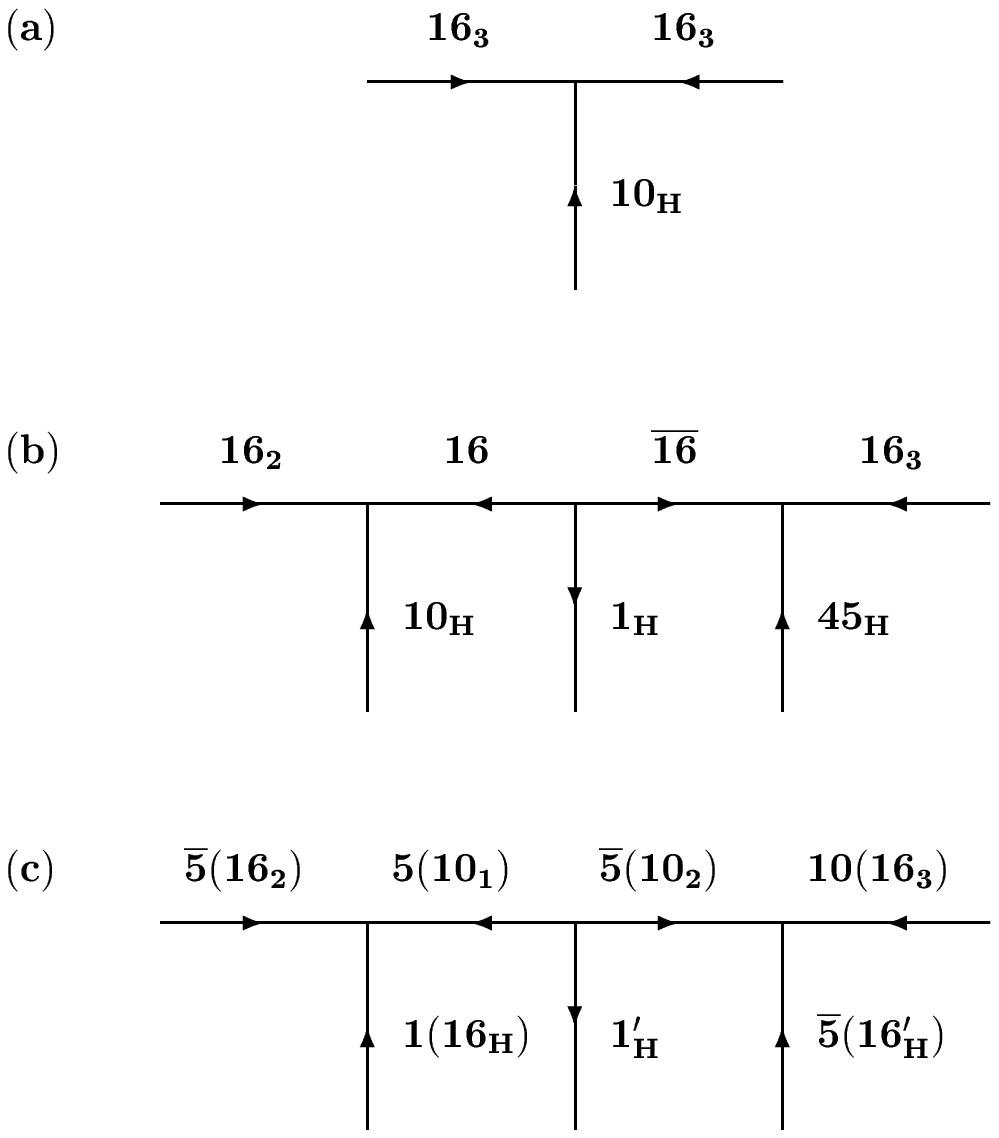}
}
\vspace {0.5in}
\caption{Diagrams that generate the 33, 23 and 32 elements in the quark and
        lepton Dirac mass matrices shown in Eqs. (\ref{eq:Dmatrices}).
        (a)~The 33 elements denoted ``1". (b) Antisymmetric contributions 
	denoted ``$\epsilon$,'' to the 23 and 32 elements where the VEV of the 
	${\bf 45}_H$ appears in the $B - L$ direction.
        (c) Asymmetric contributions to the 23 and 32 elements denoted 
	``$\sigma$" appearing in the down quark and charged lepton mass 
	matrices arise from this diagram. They do not appear in the up quark 
	and Dirac neutrino mass matrices due to the $SU(5)$ structure of the 
	fields explicitly indicated in the diagram.}
\end{figure}
\newpage
\noindent
\begin{figure}
\vspace*{0.25in}
\noindent
\centerline{
\epsfxsize=1\hsize
\epsfbox{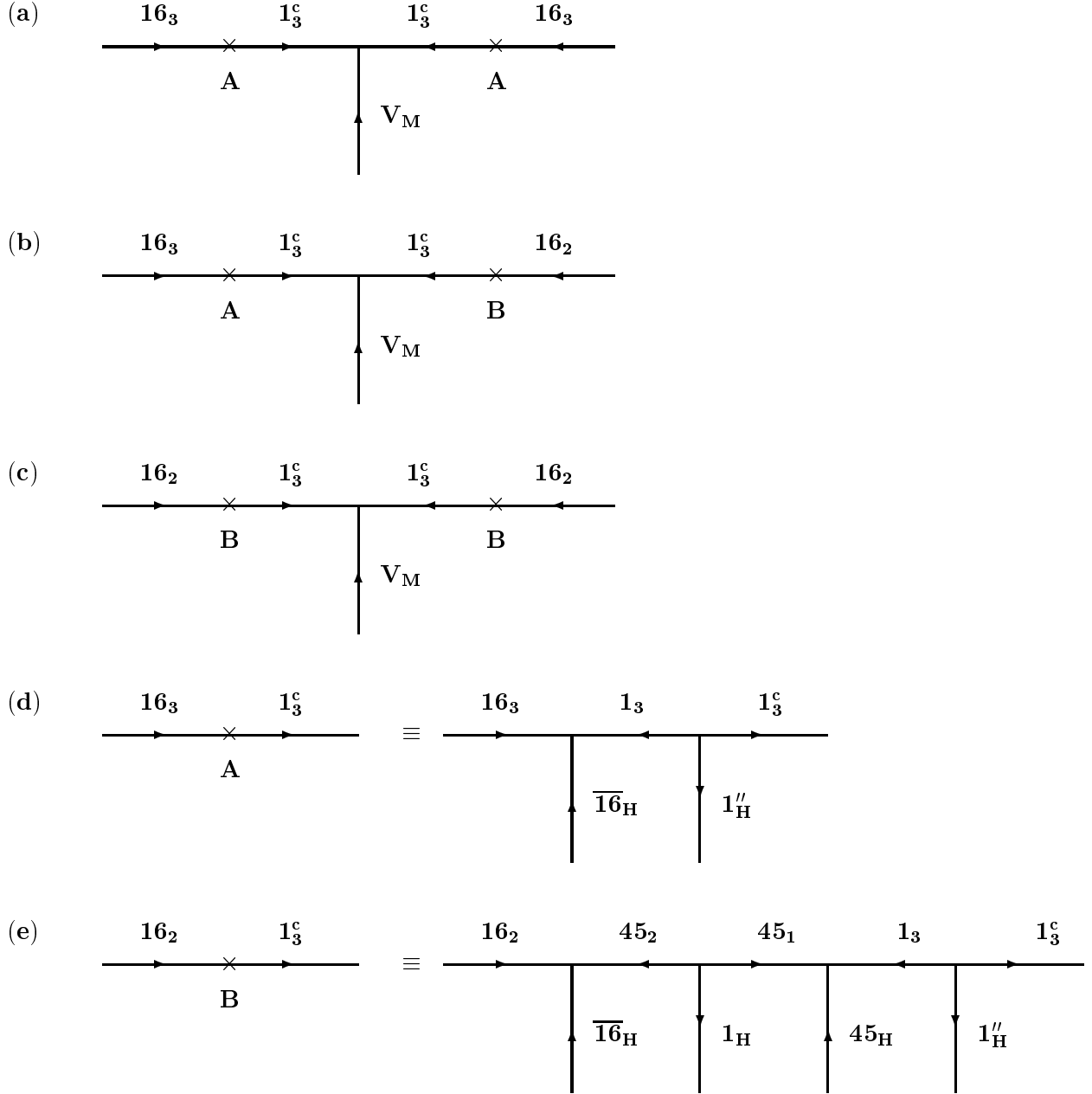}
}
\vspace{3.75in}
\caption{(a), (b) and (c), respectively, show the diagrams leading to 
	$(M_R)_{33},\ (M_R)_{32} = (M_R)_{23}$ and $(M_R)_{22}$.  Note that
	these diagrams all come from the same vertex $(1^c_3 1^c_3)V_M$
	and so lead to an exact factorized or geometrical form where
	$(M_R)_{33} (M_R)_{22} = (M_R)_{32}(M_R)_{23}$.  The insertions
	denoted ``A'' and ``B'' are defined in (d) and (e).  The ratio 
	$B/A$ is proportional to $\langle {\bf 45}_H \rangle/\langle 
	{\bf 1''}_H\rangle$ and so is of order $-\epsilon = N_{23}/N_{33}$, 
	as can be seen by inspecting Fig. 1(a) and (b).}
\end{figure}
\end{document}